\documentclass{webofc}
\usepackage[varg]{txfonts}
\usepackage{amsmath}
\usepackage{mathtools}
\usepackage{dsfont}
\usepackage{graphicx}
\usepackage{subcaption}
\usepackage[tableposition=top]{caption}
\usepackage{svg}

\begin{document}
\title{A Semi-Parametric Approach to Fitting Gas Pressure Profiles of Galaxy Clusters}

%
%

\author{\lastname{K. Wang}\inst{1}\fnsep\thanks{\email{kang.wang@vuw.ac.nz
    }} \and
        \lastname{Y. Perrott}\inst{1} \and
        \lastname{R. Arnold}\inst{1}
        \and \lastname{D. Huijser}\inst{1}
}

\institute{Victoria University of Wellington, Wellington 6140, New Zealand 
          }

\abstract{
  This study focuses on modelling galaxy cluster gas profiles via a semi-parametric nodal approach. While traditional methods like the generalised Navarro–Frenk–White (gNFW) often encounter parameter degeneracy, our flexible node-based method precisely defines a cluster gas pressure profile. Using \emph{Planck} space telescope data from the Coma region, our model, focused on the pressure-radius relationship, showcases enhanced flexibility over the gNFW. Bayesian analyses indicated an optimal five-node structure for the Coma cluster pressure profile. 
}
\maketitle
\section{Introduction}
\label{intro}
To ascertain the physical properties of galaxy clusters, it is usually necessary to perform parametric modelling, such as constructing the gas pressure profile via the gNFW \cite{nagai2007effects}:
\begin{equation}\label{eq:NFW}
	P(r)=\frac{P_{e i}}{\left(\frac{r}{r_s}\right)^\gamma\left[1+\left(\frac{r}{r_s}\right)^\alpha\right]^{(\beta-\gamma) / \alpha}}
\end{equation}
where $P_{ei}$ is a pressure normalisation factor, $r$ is radius, $r_s$ is a characteristic scaling radius and $\gamma$, $\alpha$ and $\beta$ are shape parameters. 
While the gNFW model is effective and provides good fits for clusters with diverse characteristics, it has limitations in terms of controlling parameters for the profile shape. During fitting, the scale parameters $\alpha$, $\beta$, and $\gamma $ are often degenerate, making it challenging to produce well fitted models. Degeneracy implies that different sets of parameter values can produce models equally consistent with the data.  

\cite{olamaie2018free} moved away from the traditional approach of assuming a specific parametric form for cluster properties and introduced a more flexible node-based model. This is achieved by representing  the radial profile of the  cluster  using piecewise function defined by a set of control points also called "nodes". The number of nodes, their positions and amplitudes are allowed to vary and are inferred from the data in a Bayesian approach using model selection and parameter estimation. 

In this study, we follow and further develop the method introduced by \cite{olamaie2018free} to fit \emph{Planck} data on the Coma cluster, avoiding a restrictive parameterisation and instead allowing a flexible form for the pressure-radius relationship. 
Our study diverges from \cite{olamaie2018free} in the treatment of position parameters, where we utilise Dirichlet priors to mitigate node disorder, as opposed to the uniform priors they applied (as mentioned in Section 2.3.1). While piecewise functions have been used to reconstruct pressure profiles in other work, such as \cite{romero2018multi}, where these functions rely on node positions that are fixed and must be predetermined. Our method allows the data to define where the best node positions are, and how many there should be, to best reconstruct the profile within the limitations of the resolution and sensitivity of the data. Furthermore, we intend to implement Reversible-Jump MCMC, which is efficient to explore parameter spaces and infer node counts in a singular computational execution, thereby circumventing the necessity for multiple runs and comparisons of Bayesian factors.

\section{Methods}
Our process starts with the development of a spherically symmetric 3D gas pressure model based on a multi-node approach. We then project the 3D data into a 2D perspective on the plane of the sky. Following this, the 2D model is convolved with a Gaussian Point Spread Function (PSF) to model instrument-induced blur. The model is then compared with the data. Incorporating prior beliefs, Bayesian inference is used to derive the posterior distribution of the parameters which define the model.

\subsection{Data}

\begin{figure}[h]
  \begin{minipage}[t]{0.48\linewidth}  
    \centering
    \includegraphics[scale=0.45]{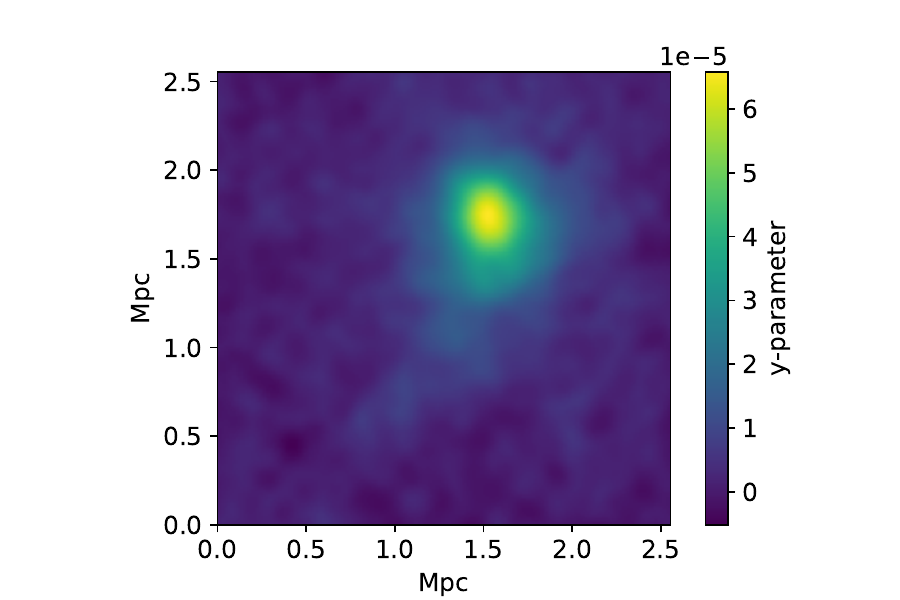}
    \caption{The tSZ signal of Coma cluster from \emph{Planck} Legacy Archive \cite{aghanim2016planck}. }
    \label{fig:Coma_cluster}
  \end{minipage}%
  \hspace{0.04\linewidth}  
  \begin{minipage}[t]{0.48\linewidth}  
    \centering
    \includegraphics[scale=0.45]{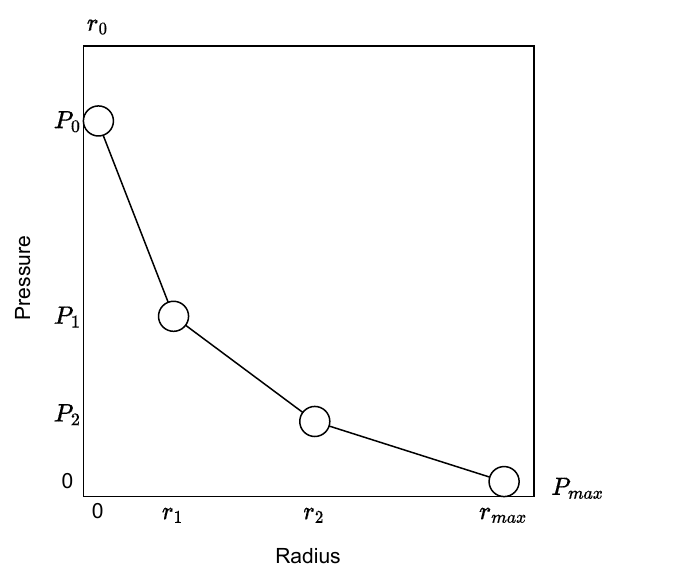}
    \caption{This is a linearly interpolated nodal representation with 4 nodes. $P_0$ means the gas pressure at a distance $r_0$ from the galaxy cluster radius. The pressure of the last node $P_{N-1}$ is fixed to $0$}
    \label{fig:node}
  \end{minipage}
  \vspace{-1.2cm}
\end{figure}

The  thermal Sunyaev-Zeldovich (tSZ) effect provides a clear measure of the accumulated thermal gas pressure along a given observational line of sight, as documented in previous studies  (e.g. \cite{birkinshaw1999sunyaev}). We obtained an image of the Compton $y$-parameter in the region of the Coma cluster (Figure \ref{fig:Coma_cluster}) from \emph{Planck} data, characterised by sharp signal clarity and an angular resolution of 10 arcmin  \cite{aghanim2016planck}. The data show a distinct and well-defined cluster with a high signal-to-noise ratio.  There is minimal interference from potential distortions such as compact sources or lingering diffuse galactic emissions—a finding consistent with Planck collaboration's examination of the Coma cluster \cite{ade2014planck}. All raw data used, including the component-separated map from \cite{aghanim2016planck} via MILCA, is available on the \emph{Planck} Legacy Archive. The map's PSF corresponds to an FWHM of $10'$, and its noise level is quantified at $2.3 \times 10^{-6}$ \cite{spergel2015planck}.

\subsection{Line-of-sight}
 The gas pressure, when integrated over the line-of-sight $\ell$, is converted into the SZ signal strength, denoted as the Compton $y$-parameter:
\begin{equation}
y(r_{\text{proj}}) = \int \frac{\sigma_T}{m_e c^2} P(r) d \ell,
\end{equation}
where $\sigma_T$ represents the Thomson cross-section, $m_e$ is the electron mass, and $c$ is the speed of light.  
Each pixel on our sky grid requires calculation of this integral, with 
$r_{\text{proj}}$ representing the projected radius on the sky relative to the cluster centre. Adopting $r =\sqrt{r_{proj}^2+\ell^2}$, then: 
\begin{equation}\label{eq:y}
y(r_{proj}) = \frac{ \sigma_T }{m_e c^2} \int^ {\infty}_{-\infty} P(\sqrt{r_{proj}^2+\ell^2}) d\ell .
\end{equation}

\subsection{Semi-parametric nodal model}
We  specify a multi-nodal model which is a piecewise linear model for the pressure profile following \cite{olamaie2018free}. In this initial study, we assume the special case of a spherical cluster. We describe the relationship between the galaxy cluster radius and the gas pressure by constructing a piecewise specification of the cluster pressure profile $P(r)$. Each node $(P_i, r_i)$ is a control point for this piecewise function. The first node radius $r_{0}$ and outermost node pressure $P_{N-1}$ are fixed, in order to stabilise the estimation process. It turns out that continuous free form reconstructions of the profile $P(r)$ can be obtained, each node between first and last nodes can move vary both $r_i$ and $P_i$. For example, we express the relationship between cluster gas pressure and radius using a piecewise function:
\begin{equation} \label{eq:47}
	P\left( r\right) = \sum_{i=0}^{N-2} \left( \dfrac {P_{i}-P_{i+1}}{r_{i}-r_{i+1}}\left( r-r_{i}\right) +P_{i}\right) \mathds{1}_{ \left[ r_{i}\leq r\leq r_{i+1}\right]}, ~~~ i = 0,1,..., N - 2
\end{equation}
where $N$ is the number of total nodes. For clarity, we refer to the last node $(P_{N-1}, r_{N-1})$ as $(P_{N-1}, r_{N-1})$.  The indicator function is  $\mathds{1}$ and defined as: 
\begin{equation}\label{eq:indx}
\mathds{1}_{ \left[ r_{i}\leq r\leq r_{i + 1}\right]}:={\begin{cases}1~ &{\text{ if }}~r\in (r_{i}, ~r_{i + 1}),\\0~&{\text{ if }}~ r\notin (r_{i},~ r_{i + 1}).\end{cases}}  ~~~ i = 0,1,..., N - 2.
\end{equation}
The model is illustrated in figure 2, with an example pressure profile with $N=4$ nodes.

\vspace{-0.1cm}

\subsubsection{Prior}

In our Bayesian analysis, our sampling parameters are the position of the centre of the galaxy cluster in 2D projection, the radius of each node from the centre, the gas pressure on the node and the standard deviation of the signal noise. The parameters are:
\begin{equation}
\Theta  \equiv (X_0, Y_0, \sigma, P_0, P_1,..., P_{N-1}, r_1,..., r_{N-1}).
\end{equation}
where $(X_0, Y_0)$ is the central position, and $\sigma$ is the noise parameter. We require the radius parameter to have the following properties: $r_0<r_1<r_2...<r_{N-1}$. The position priors $r_i ~  \text{for}~ i \geq 1$ follow a Dirichlet-relative distribution. For an explanation of Dirichlet-relative distribution, see the end of this section.  The pressure of the first node $P_0$ has a truncated exponential distribution with $\lambda = \frac{1}{1\times10^{-19}} \rm{\frac{Mpc \cdot s^2}{M_\odot}}$ because our prior experience believes that the mean of $P_0$ is equal to ${1\times10^{-19} \rm{\frac{M_\odot}{Mpc \cdot s^2}}}$  and in the range of $1\times10^{-24} \leq P_0/ \rm{\frac{M_\odot}{Mpc \cdot s^2}} \leq 5\times10^{-19}$. The remaining node pressures have  uniform priors within $0 \leq P_i /\rm{\frac{M_\odot}{Mpc \cdot s^2}} \leq 5\times10^{-20}$. We fix $r_{0}  $ and $P_{N-1}$ to be $ 0~ \rm{Mpc} $ and $0~ \rm{\frac{M_\odot}{Mpc \cdot s^2}}$ respectively. The position parameters of the galaxy cluster centre are also uniform priors covering all pixels of the image. Finally, we assume that the signal noise $\sigma $ has a uniform prior within $0 \leq  \sigma \leq 3\times10^{-6}$. Since our model employs a radially symmetric mean model, which may not be exactly representative of the true cluster shape (see Fig. 1), there will be departures of the data from the model which are not due to noise (intrinsic scatter).  Allowing the noise parameter to vary, rather than fixing it to the instrumental noise, allows us to include this intrinsic scatter in the likelihood calculation.

The MultiNest algorithm, first introduced by \cite{feroz2009multinest}, is a Bayesian inference tool adept at dealing with multimodal posteriors and calculating the evidence. In practice, it utilises a D-dimensional unit hypercube  where every parameter value varies between 0 and 1, with samples being uniformly random drawn within these parameters. When we set the prior, we need to use the inverse cumulative distribution function (CDF) method to ensure the parameters follow the desired distribution. The Dirichlet-relative prior is constructed as follows:
\begin{align} 
&U_{i}\sim U\left( 0,1\right) ~\text{for} ~ i = 1,2,...,N \nonumber \\ \nonumber
&x_{i}=-\ln U_{i} \sim  \rm{Exp} ~ (1)  \\ \nonumber
&d_{i}=\dfrac {x_{i}}{\sum _{j}x_{j}} \sim  \rm{Dir} ~(\underbrace{1,1,..,1}_{N ~ \text{times}}) \\ \nonumber
&w_{i}=\sum _{k=1}^{i}d_{k}=\dfrac {\sum ^{i}_{k=1}x_{k}}{\sum ^{N}_{j=1}x_{j}}=\dfrac {\sum ^{i}_{k=1}-\ln U_{k}}{\sum ^{N}_{j=1}-\ln U_{j}}\\ 
&r_{i}=r_{0}+w_{i}\left( r_{N-1 }-r_{0}\right). 
\end{align}
We require the parameters $r_i$ in order and that $r_{N-1}$ is the maximum radius of the cluster. Then, the vector $(d_1,...,d_N)$ follows a Dirichlet (1,...,1) distribution which can be regarded as normlised sum of independent Exp (1) random variables.

\subsubsection{Likelihood Function}
Next, we define the likelihood. The standard deviation of the Gaussian noise $\sigma$ is a parameter, and the likelihood is:
\begin{equation}\Pr(\mathbf{D}|\Theta,\mathbf{M}) = (2\pi)^{-m/2} \exp \left [-\chi^2/2 \right ] \sigma^{-m}\end{equation}
where $\mathbf{D} = \{ D_k \}^{m}_{k=1} $ is the $y$-parameter in each pixel and ${M}_k = y[{x_k}, {y_k} | \Theta ]$ is the model in each pixel $k$. According to Eq \ref{eq:y}, we have $y[r_{proj} = \sqrt{(x_k-x_0)^2+(y_k-y_0)^2})| \Theta]$. The number of pixels is $m$. The error, $\sigma$ is assumed to be the same for each pixel. The chi-squared statistic is $\chi^2$: \begin{equation}\chi^2 = \sum_{k=1}^m \left ( \frac{{D}_k - {M}_k}{\sigma} \right )^2.\end{equation} MultiNest works with the natural log of the likelihood:

\begin{equation}\ln \Pr(\mathbf{D}|\Theta,\mathbf{M}) = -\frac{m}{2} \ln (2\pi) - \frac{1}{2} \chi^2 - m \ln \sigma.\end{equation}

\section{Results}

\begin{figure}[h]
  \begin{minipage}[t]{0.48\linewidth}  
    \centering
    \includegraphics[scale=0.45]{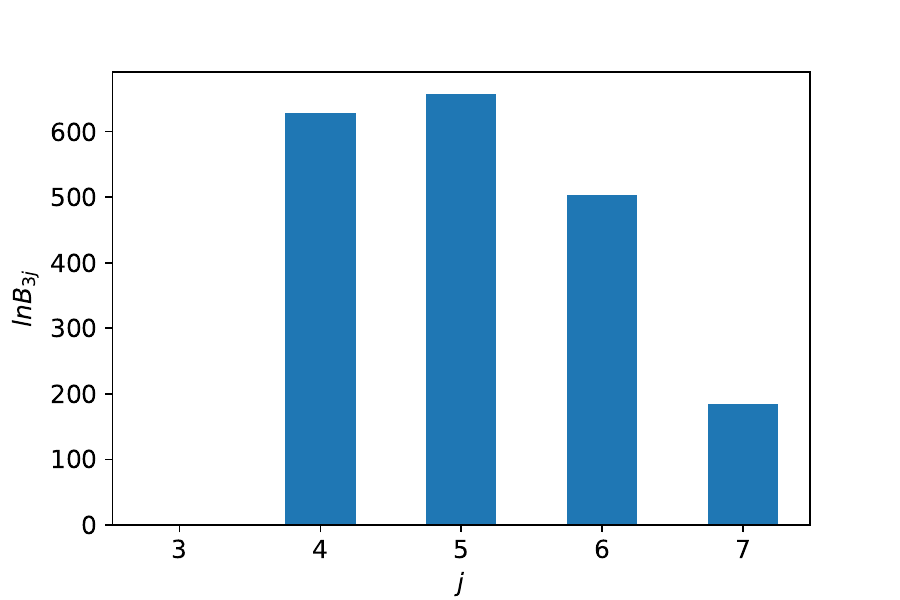}
    \caption{Histogram of log Bayes factors $\ln \mathcal{B}_{3j}$ relative to the $j$ = 3 node model, for the fit to the \emph{Planck} Coma data. }
    \label{fig:Bij}
  \end{minipage}%
  \hspace{0.04\linewidth}  
  \begin{minipage}[t]{0.48\linewidth}  
    \centering
    \includegraphics[scale=0.45]{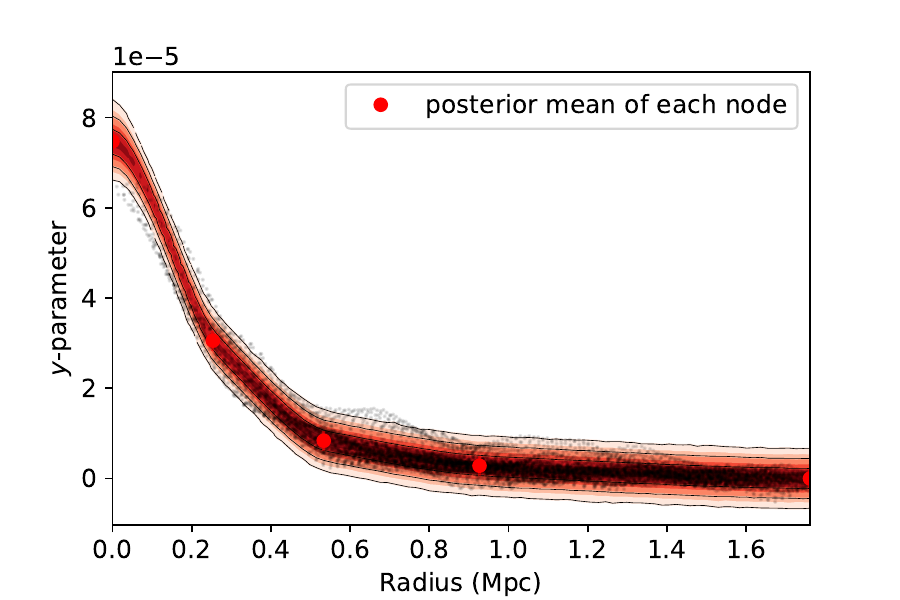}
    \caption{Posterior predictive distribution plot in 5 node model. The red contour plot represents the probability of the next new observation point given the data and the model, taking into account the noise as well as the uncertainty on the model fit. Black dots are observed data.}
    \label{fig:straight4line4}
  \end{minipage}
  \vspace{-0.35cm}
\end{figure}

We compare the models with different numbers of nodes through a histogram of Bayes factors in Figure \ref{fig:Bij}. The Bayes factor \(\mathcal{B}_{ij} \equiv \frac{Z_j}{Z_i}\) is a ratio representing the evidence comparison between model \(j\) and model \(i\).  It is commonly utilized for model comparison, as established by Jeffreys \cite{jeffreys1961}.
  Bayes factors $\ln \mathcal{B}_{3j}$  is the comparison of a model with $j$ nodes to the 3 node model. When $j = 5$, the logarithm of the Bayes factor $\ln \mathcal{B}_{35}$ is maximal for the Coma cluster pressure profile. Specifically, the log Bayes factor  $\ln \mathcal{B}_{35} = 657.879$ implies substantial support for the 5 node model over the 3 node model, according to Jeffreys' scale \cite{jeffreys1961}.  Bayes factors  slowly decrease for $j>5$, this shows that even after increasing the number of free points, we did not achieve better results.

\vspace{-0.15cm}

In Figure \ref{fig:straight4line4}, the red heat map  depicts the posterior predictive distribution for the 5 node model. This diagram shows the goodness of fit of the model to the data. The black points signify the Coma cluster $y$-map pixel values as a function of radius originating from the cluster centre. As evident from the figure, our posterior predictive contour nearly encompasses the actual observational data. Compared with the gNFW method, the use of the semi-parametric nodal model can match the data more flexibly.

\begin{figure}[h]
\centering
\includegraphics[width= 4.8 in, keepaspectratio]{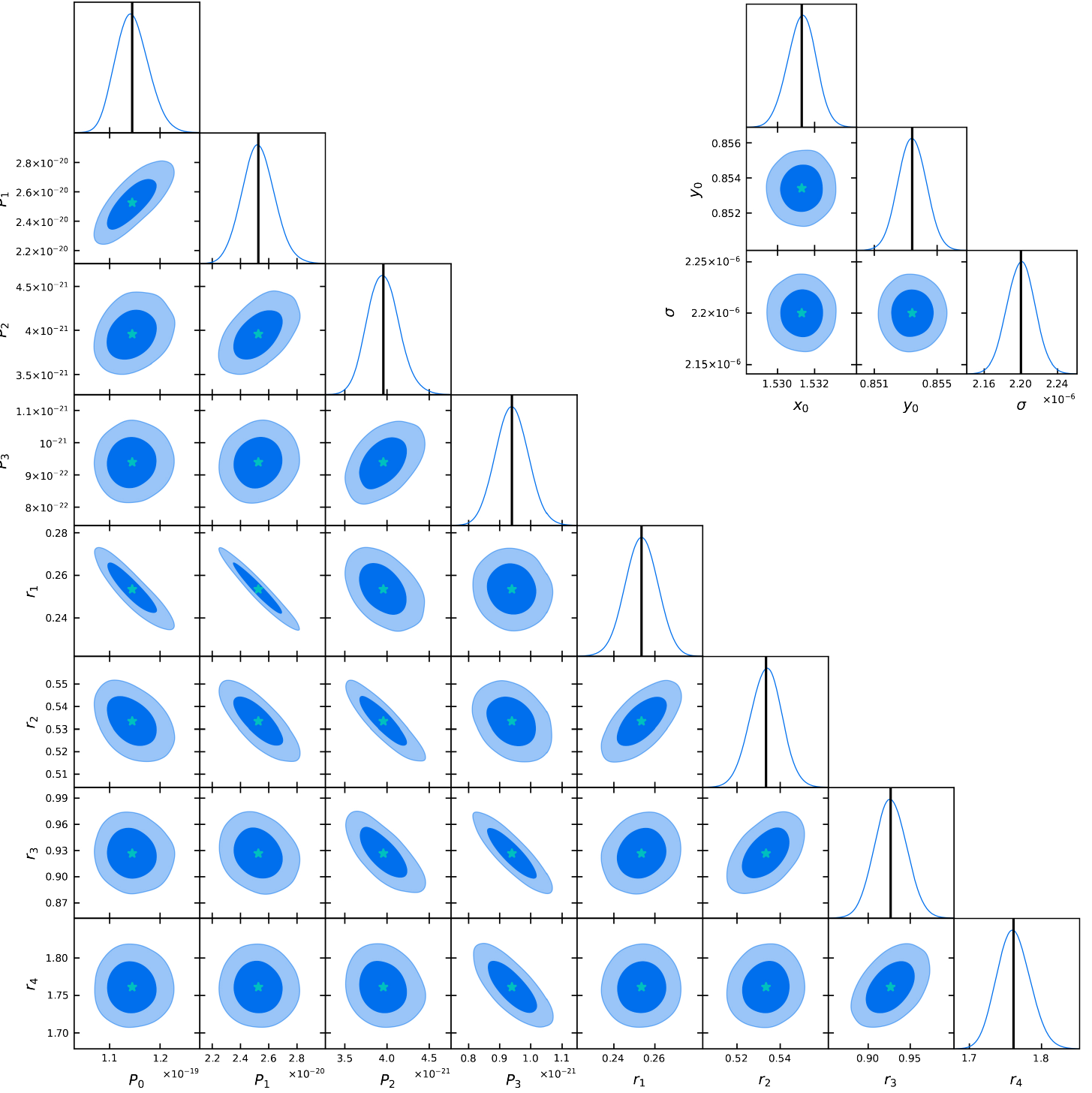}\\
\caption{
1D and 2D posterior probability distributions for the parameters of the 5  nodes model. Each contour is drawn at both 68 \% and 95 \% credible interval (blue and light blue).  The asterisk represents the mean value of the posterior distribution of each parameter.}
\label{fig:3tri}
\vspace{-0.5cm}
\end{figure}

The  corresponding triangle plot of the posterior probability distribution of the model parameters is shown in Figure \ref{fig:3tri} for the optimal Bayesian factor model $j=5$. Overall, most parameters show posterior independence, but some individual pairs of parameters show correlations. For example, $(P_1, r_1)$ and $(P_2, r_2)$ both show a strong negative correlation, which is expected, because the closer to the center of the cluster, the greater the pressure. 

\vspace{-0.5cm}

\section{Conclusion}

In this study, we conducted fitting and analysis of a semi-parametric model of the pressure profile of the Coma cluster using \emph{Planck} data. In future work, we will use powerlaw interpolation between control points. This will improve the reconstruction of the profile since in general cluster pressure profiles are expected to be closer to power-law than linear behaviour.  Although analytic solutions exist to the line-of-sight integral for power-law profiles, their computation is more expensive than the simple linear integration, increasing run-times, so our initial development work has been done using linear functions. We also plan to fit a range of clusters including clusters with low signal to noise ratios. A notable advancement in our methodology will be the amalgamation of DNest 4 and Reversible Jump MCMC \cite{brewer2018dnest4}. This integration will implement an automated trans-dimensional technique, facilitating model selection within a singular program execution instead of running the program multiple times and comparing the Bayes factor.


%
%

\end{document}